\documentclass[fleqn,twoside]{article}      
\usepackage{epsf,multicol,ifthen}
\usepackage{ujp}
\usepackage[cp1251]{inputenc}
\usepackage[english,ukrainian]{babel}
\usepackage{amstext}
\usepackage{graphicx}

\nazva{SHORT-WAVE EXCITATIONS IN NON-LOCAL GROSS-PITAEVSKII MODEL}%
\udk{532.591,538.94,539.196.3} \nazvacol{SHORT-WAVE EXCITATIONS IN NON-LOCAL GROSS-PITAEVSKII MODEL}%
\avtor{A.P. IVASHIN, Yu.M. POLUEKTOV}%
\avtorcol{A.P. IVASHIN, Yu.M. POLUEKTOV}%
\inst{A.I.~Akhiezer Institute for Theoretical Physics,\\ NSC ``Kharkov Institute of
Physics and Technology''}%
\adr{(1, Akademicheskaya Str., Kharkov UA-61108, Ukraine;\\
 e-mail: ivashin@kipt.kharkov.ua,yuripoluektov@kipt.kharkov.ua)}%

\begin{document}
\maketitl                 
\begin{multicols}{2}
\anot{%
It is shown, that a non-local form of the Gross-Pitaevskii equation allows
to describe not only the long-wave excitations, but also the short-wave ones
in the systems with Bose-condensate. At given parameter values, the
excitation spectrum mimics the Landau spectrum of quasi-particle excitations
in superfluid Helium with roton minimum. The excitation wavelength, at which
the roton minimum exists, is close to the inter-particle interaction range.
It is shown, that the existence domain of the spectrum with a roton minimum
is reduced, if one accounts for an inter-particle attraction.}%

\noindent
The Gross-Pitaevskii equation (GP) obtained in~\cite{G,P,LP}
is widely used in researches of superfluid Bose-systems
of weakly interacting particles.
This equation describes the particle dynamics in a Bose-condensate.
In particular, it gives a framework to study the condensate
oscillations and to obtain the elementary excitation spectra.
The knowledge of the latter is necessary in order
to obtain the thermodynamic functions of the system.

At a stage of derivation of the GP equation for
the condensate wave function $\Phi \left( {r,t} \right)$,
it is usually assumed that $\Phi$ varies
slowly over the distances of order $r_{0}$, which is
the range of the inter-atomic interaction potential $U\left( {r} \right)$.
This condition allows to express the GP equation in form
of a non-linear differential equation.

Unfortunately, in such a case, the imposed assumption
restricts a consideration to the long-wave excitations
in the Bose-condensate. Namely, it allows one to
derive the Bogolyubov sound (long-wave) oscillation spectrum;
nevertheless, a (physical) superfluid system can also have
some well-defined long-lived short-wave excitations
with the de Broglie wavelengths of the order
of inter-particle distance (or interaction
potential range).
E.g., in the superfluid helium the rotons are the
excitations of this type.

The purpose of this paper is to demonstrate that
the non-local form of the GP equation
describes the short-wave excitations in a system with Bose-condensate,
as well as the long-wave ones,
if the constraint on the of macroscopic wave function spacial change scale
is released.
At given values of the parameters the excitation spectrum
mimics the Landau one for the quasi-particles in the superfluid
Helium, which has the roton minimum.
We study the influence of the inter-particle interaction on the
dispersion curve in the short-wave domain.
It is shown that the inter-particle attraction
narrows down the existence domain of a stable spectrum.
In particular, it reduces the range of parameters where a
spectrum with roton minimum may exist.

\bigskip

\section*{NON-LOCAL FORM OF GROSS-PITAEVSKII EQUATION}

A system of many Bose-particles in the second quantization picture
with an account for the pairwise interactions is described by the
Hamiltonian
\begin{eqnarray}
\label{eq0} H \!\!\!&=&\!\!\!\int {d{\mathbf{r}}} \hat \psi^{+}
\left( {\mathbf{r}} \right) \left(- {\frac{{\hbar
^{2}}}{{2m}}\Delta} \right)\hat \psi \left( {\mathbf{r}} \right)
\\
&&\!\!\!\!\! +\frac{{1}}{{2}} \int\!\!\!\int
{d{\mathbf{r}}}{d{\mathbf{r}}'} \hat\psi^{+} \left( {\mathbf{r}}
\right)\hat\psi^{+} \left( {\mathbf{r}}' \right) {\kern 1pt}
U\left( {\mathbf{r} - {\mathbf{r}}'} \right)\hat\psi \left(
{\mathbf{r}}' \right) \hat\psi \left( {\mathbf{r}} \right).
\nonumber
\end{eqnarray}
The equation of motion for the field operator in the
Heisenberg picture reads
\begin{eqnarray}
\label{eq1} i\hbar \frac{{\partial \hat\psi \left( {\mathbf{r},t}
\right)}}{{\partial {\kern 1pt} t}} &=& -  {\frac{{\hbar
^{2}}}{{2m}}\Delta }  \hat\psi \left( {\mathbf{r},t} \right)
\\
&&+ \psi \left( {\mathbf{r},t} \right)\int {d{\mathbf{r}}'} {\kern
1pt} U\left( {\mathbf{r} - {\mathbf{r}}'} \right)\left| {\hat\psi
\left( {{\mathbf{r}}',t} \right)} \right|^{2}, \nonumber
\end{eqnarray}

Due to a large number of particles in the condensate,
one may consider the operator
 $\hat \psi(\mathbf{r},t)$
 in eq.~(\ref{eq1})
 as a non-operator function
  $\psi \left( {\mathbf{r},t} \right)$.
 Further, it is convenient to define a function
  $\Phi \left( {\mathbf{r},t} \right)$:
\begin{eqnarray}
\psi \left( {\mathbf{r},t} \right) = \Phi \left( {\mathbf{r},t}
\right) \exp\left( { - i\mu {\kern 1pt} {\kern 1pt} t} \right).
\end{eqnarray}
Then the equation reads
\begin{eqnarray}
\label{eq2} i\hbar \frac{{\partial \Phi \left( {r,t}
\right)}}{{\partial {\kern 1pt} t}} &=& - \left( {\frac{{\hbar
^{2}}}{{2m}}\Delta + \mu}  \right)\Phi \left( {r,t} \right)
\\
&&+ \Phi \left( {r,t} \right)\int {d{r}'} {\kern 1pt} U\left( {r -
{r}'} \right)\left| {\Phi \left( {{r}',t} \right)} \right|^{2},
\nonumber
\end{eqnarray}
\noindent where $\mu$ is the chemical potential
which value is determined by the equilibrium of the system
\begin{eqnarray}
\label{eq4a} \mu=\int {d{\mathbf{r}}'} {\kern 1pt} U\left(
{\mathbf{r} - {\mathbf{r}}'} \right)\left| {\Phi_{0} \left(
{{\mathbf{r}}'} \right)} \right|^{2},
\end{eqnarray}
\noindent with $\Phi _{0} \left( {\mathbf{r}} \right)$
 the equilibrium value of the condensate wave function.
Taking into account that the equilibrium condensate density in
absence of external fields $n_{0} = \left| {\Phi _{0} \left(
{\mathbf{r}} \right)} \right|^{2}$ does not depend on the
coordinate, one can express eq.~(\ref{eq2}) as
\begin{eqnarray}
\label{eq3} i\hbar \frac{{\partial \Phi \left( {\mathbf{r},t}
\right)}}{{\partial {\kern 1pt} t}} &=& - \left( {\frac{{\hbar
^{2}}}{{2m}}\Delta + n_{0} U_{0}}  \right)\Phi \left(
{\mathbf{r},t} \right)
\\
&&+ \Phi \left( {\mathbf{r},t} \right)\int {d{\mathbf{r}}'} {\kern
1pt} U\left( {\mathbf{r} - {\mathbf{r}}'} \right)\left| {\Phi
\left( {{\mathbf{r}}',t} \right)} \right|^{2}, \nonumber
\end{eqnarray}
\noindent where
 $U_{0}=\int {d\mathbf{r}'} {\kern 1pt}
 U\left( {\mathbf{r} - \mathbf{r}'} \right)$.
 If one is interested only in the long-wave perturbations,
 and assumes that the characteristic spatial scale of the condensate
 wave function is much greater than the inter-particle
 interaction potential range,
 then the square of the wave function can be taken out of the
 integral in the right hand side of eq.~(\ref{eq3}).
 Thus, the GP equation takes the form
\begin{eqnarray}
\label{eq4} i\hbar \frac{{\partial \Phi \left( {\mathbf{r},t}
\right)}}{{\partial {\kern 1pt} t}} &=& - \left( {\frac{{\hbar
^{2}}}{{2m}}\Delta + n_{0} U_{0}}  \right)\Phi \left(
{\mathbf{r},t} \right)
\\
&&+ U_{0} \Phi \left( {\mathbf{r},t} \right) {\kern 1pt} \left|
{\Phi \left( {{\mathbf{r}},t} \right)} \right|^{2}. \nonumber
\end{eqnarray}
This is the usual way to apply the GP equation.
Small oscillations described by eq.~(\ref{eq4})
have the dispersion law
\begin{equation}
\label{eq5} \omega ^{2} = \frac{{U_{0} n_{0} k^{2}}}{{m}} +
\frac{{\hbar ^{2}k^{4}}}{{4m^{2}}},
\end{equation}
\noindent where $\omega$, $k$ are the frequency and the
wave number.
The dispersion law~(\ref{eq5}) was derived by Bogolyubov~\cite{B}
in a different approach with the use of the second quantization method.
The excitation energy due to eq.~(\ref{eq5})
linearly depends on the wave number for small $k$
and is monotone increasing with $k$,
 approaching the free particle dispersion law.

One should stress that spectrum~(\ref{eq5})
is valid only for the long-wave oscillations due to
the approximation used for derivation of
eq.~(\ref{eq4}).

Meanwhile, the derivation of the GP equation in form~(\ref{eq3})
does not assume that the excitations are of long-wave type.
Thus, eq.~(\ref{eq3}) is also valid for description of
the excitations which wave length is of order of the inter-particle
interaction. We call further eq.~(\ref{eq3}) the
non-local Gross-Pitaevskii equation (NGP).
This equation can be applied in order to study
the dispersion law of the collective excitations of a Bose-condensate
in a short-wave domain of spectrum
and to analyze the spatial behavior of the condensate wave function
at the distances of order of the inter-atomic potential range.

\bigskip

\section*{HYDRODYNAMIC FORM OF NON-LOCAL GROSS-PITAEVSKII EQUATION}

Equation~(\ref{eq3}) can be rewritten in the form of hydrodynamic
equations, if the condensate wave function is taken as
$\Phi = \eta {\kern 1pt} e^{i\varphi} $,
where
$\varphi ,\,{\kern 1pt} \eta $ are the phase and the absolute value
of the condensate wave function and
$\eta ^{2} = n$ is the particle number density in the condensate.

Taking the above into account, eq.~(\ref{eq3}) leads to
\begin{eqnarray}
\label{eq6}
\frac {\partial n}{\partial t} +\nabla (n
\mathbf{v})&=&0,
\\
\label{eq7}
\frac{\hbar}{m}\frac {\partial \varphi}{\partial t}
&=&-\frac{
\mathbf{v}^2}{2}+\frac{\hbar^2}{4m^2}\left(\frac{\Delta n}{n}
-\frac{(\nabla n)^2}{2n^2}\right)
\\
&&
+\frac{n_{0}U_{0}}{m}-\frac{1}{m}\int d{\mathbf{r}}' U\left(
{\mathbf{r} - {\mathbf{r}}'} \right)n({\mathbf{r}}'), \nonumber
\end{eqnarray}

\noindent where the notion of
superfluid velocity is introduced as
\begin{eqnarray}
\mathbf{v} &=& \left( {{{\hbar} \mathord{\left/ {\vphantom {{\hbar}
{m}}} \right. \kern-\nulldelimiterspace} {m}}} \right)\nabla
\varphi .
\end{eqnarray}
Eq.~(\ref{eq6}) is the continuity equation,
and eq.~(\ref{eq7}) is analogous to the Josephson equation in the
theory of superconductivity.
Taking a gradient of both sides of eq.~(\ref{eq7})
and taking into account the definition for the velocity,
one gets
\begin{eqnarray}
\label{eq8} \frac {\partial \mathbf{v}}{\partial t} +(
\mathbf{v}\nabla) \mathbf{v}=\nabla w,
\end{eqnarray}
\noindent where
\begin{eqnarray*}
\label{eq11a} w=\frac{\hbar^2}{2m}\frac{\Delta
\eta}{\eta}-\frac{1}{m}\int {d{\mathbf{r}}'} U\left( {\mathbf{r} -
{\mathbf{r}}'} \right)n({\mathbf{r}}').
\end{eqnarray*}

In a linear approximation, the right hand side of eq.~(\ref{eq8})
can be expressed in form
$\nabla w = - {{\nabla p} \mathord{\left/
{\vphantom {{\nabla p} {mn_{0}} }} \right.
\kern-\nulldelimiterspace} {mn_{0}} }$,
where the pressure and the density are related in
a non-local way

\begin{eqnarray}
\label{eq12b} p=-\frac{\hbar^2}{2}\eta_{0}\Delta \eta +n_{0}\int
{d{\mathbf{r}}'} U\left( {\mathbf{r} - {\mathbf{r}}'}
\right)n({\mathbf{r}}'),
\end{eqnarray}

Non-local relation between the pressure and the density
was used in the phenomenological approach of~\cite{A}
(neglecting the quantum term) in order to describe
the rotons in the superfluid Helium.
From our consideration it follows that the non-locality kernel,
which relates the pressure and the density in a phenomenological
approach, is determined by the inter-particle interaction potential.

\bigskip

\section*{SMALL OSCILLATION SPECTRUM WITH NON-LOCAL EFFECTS}

The non-local GP equation leads to the following dispersion
law for small oscillations
\begin{eqnarray}
\label{eq7b} \omega ^{2} = \frac{{U_{k} n_{0} k^{2}}}{{m}} +
\frac{{\hbar ^{2}k^{4}}}{{4m^{2}}},
\end{eqnarray}
\noindent which is different from~(\ref{eq5}) due to appearance of
a potential Fourier component $U_{k} = \int {d\mathbf{r}{\kern
1pt}}  U\left( {\mathbf{r}} \right)e^{i{\kern 1pt} kr}$ (dependent
on the wave vector)
 instead of $U_{0}$.

Usually, at short distances, a strong repulsion occurs,
thus the interaction potential grows rapidly.
It leads to a divergence of the Fourier component
$U_{k}$.
In order to
overcome this obstacle
one may suppose that the potential remains finite
at short distances~\cite{EP}.
As the simplest case, we will use
the "`semitransparent sphere model"' potential
\begin{eqnarray}
\label{eq7c} U\left( {\mathbf{r}} \right) = \left\{
\begin{array}{ll}
U_m, & r \le a, \\
0,   & r > a,
\end{array}
\right.
\end{eqnarray}
\noindent and consider $U_{m}$ and $a$ as two independent
parameters of the dimension of energy and length, correspondingly.
In this case the Bose-condensate excitation spectrum
depends on the only dimensionless parameter
\begin{equation}
  \label{eq9}
  \xi = \frac{{\hbar ^{2}}}{{8\pi\ n_{0}\ a^5\ m\ U_m}},
\end{equation}
\noindent and can be written in form
 \begin{equation}
  \label{eq9b}
  \omega ^{2} = \omega
_{0}^{2} f\left( {z} \right),
\end{equation}
\noindent
where
 \begin{eqnarray}
  \label{eq10}
  f(z) &=& z^{-1}(\sin z - z \cos z)+\xi z^4,
  \\
  \label{eq10b}
  \omega_{0}^{2} &=& {4\pi\ n_{0}\ a\ U_{m}}/{m}
  \\
  \label{eq10c}
  z&=&ka.
 \end{eqnarray}
Notice that the dimensionless parameter
$\xi$~(\ref{eq9})
strongly depends on the potential range
$\xi \propto a^{-5}$,
rapidly growing with decreasing~$a$.

The dependence of condensate excitation frequency on the wave number
is shown in Fig.1.
At parameter values $\xi > \xi_{h} = 3.4 \cdot 10^{-3}$,
the dispersion curve does not have local minima
and the energy monotone increases with wave number
(upper curve in Fig.1).

\bigskip
\begin{center} \noindent
\epsfxsize=0.8\columnwidth\epsffile{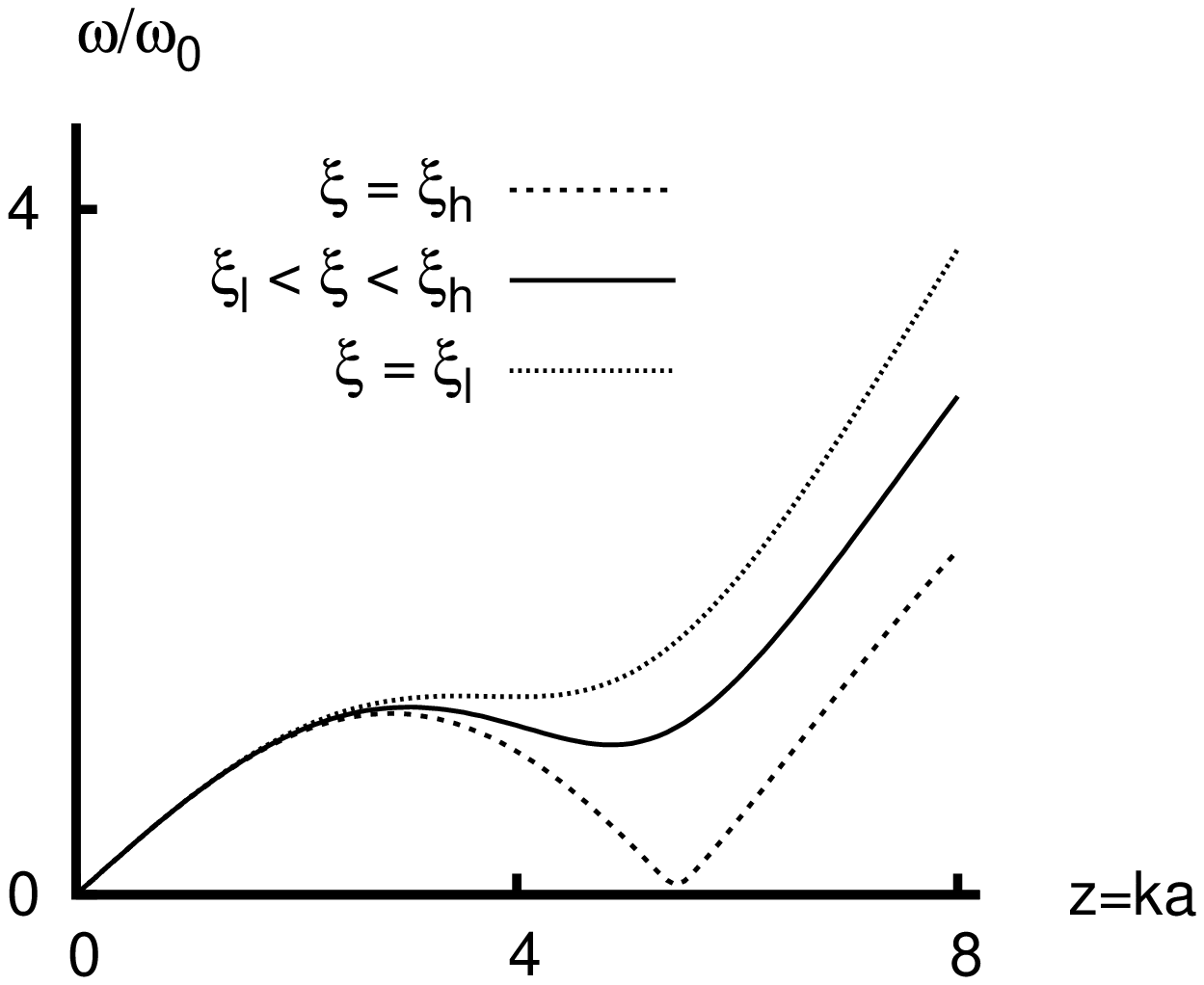}
\end{center}

\vskip-3mm\noindent{\footnotesize
Fig.1. Condensate excitation spectrum
for the "`semitransparent sphere model"' interaction;
$\xi_{h} = 3.4 \cdot 10^{-3}$,
$\xi_{l} = 0.92 \cdot 10^{-3}$
                   }%
\vskip15pt
\noindent
At $\xi_{l} < \xi < \xi_{h}$ the dispersion curve acquires
a roton minimum (middle curve in Fig.1)
and its depth increases with decreasing $\xi$.
At the value $\xi = \xi_{l} = 0.92 \cdot 10^{-3}$,
the excitation energy minimum equals to zero
(lower curve in Fig.1); it means an instability
of a many-particle system for given parameters of
the interaction potential.
Thus, in the range $\xi_{l} < \xi < \xi_{h}$
the excitation spectrum mimics that of superfluid Helium
with a roton minimum~\cite{H}.

The roton minimum position $z_{0}$ is determined by
the equation ${f}'\left( {z_{0}}  \right) = 0$.
Dependence of $z_{0}$ on the parameter $\xi$ is shown in Fig.2.
With increasing  $\xi$  the roton minimum position
gets shifted toward smaller wave numbers,
changing from its maximum $z_{max} = 5.45$
(for $\xi_{l} = 0.92 \cdot 10^{-3}$) to minimum
$z_{min} = 3.72$
(for $\xi _{h} = 3.4 \cdot 10^{-3}$).
It follows that the excitation wave length $\lambda$,
at the dispersion curve minimum
lies in the range
\begin{equation}
\label{eq:range:lambda}
1,15{\kern 1pt} a < \lambda < 1,69{\kern 1pt} a.
\end{equation}
Thus one may conclude that the existence of a dispersion curve minimum
is caused by the finiteness of the inter-particle interaction range
and is not connected with any specific superfluid properties of system.
It is noteworthy that in Helium
the repulsive range of the inter-atomic interaction potential
may be estimated as $a = 2.56\,$\r{A}
and the roton wave length is $\lambda_{r} = 3.3\,$\r{A}
such that the relation $\lambda_{r} = 1.28{\kern 1pt} a$
fits into the interval~(\ref{eq:range:lambda}).

Dependence of the roton gap
$\Delta = \omega _{0} \sqrt {f\left( {z_{0}} \right)}$
of the spectrum~(\ref{eq9b}),~(\ref{eq10})
on the parameter $\xi$ is monotone;
Fig.3 shows that $\Delta \left/\omega _{0} \right.$
 is zero at $\xi = \xi_{l}$
 and achieves its maximum
${{\Delta _{max}}  \mathord{\left/ {\vphantom {{\Delta
_{max}} {\omega _{0} = 1.17}}} \right. \kern-\nulldelimiterspace}
{\omega _{0} = 1.17}}$
at $\xi = \xi_{h}$.

\bigskip
\begin{center} \noindent
\epsfxsize=0.8\columnwidth\epsffile{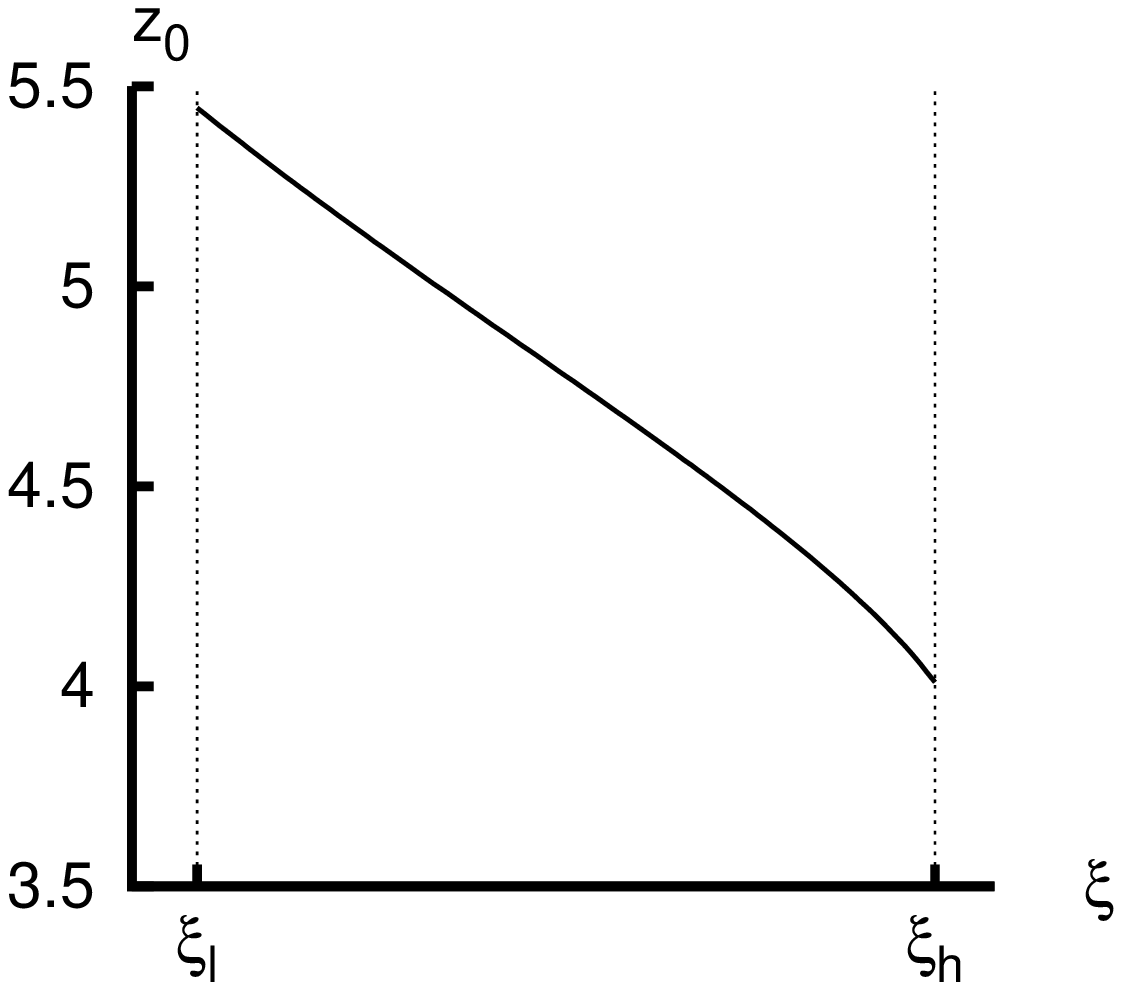}
\end{center}
\vskip-3mm\noindent{\footnotesize
Fig.2. Roton minimum position $z_{0}$ as a function
of parameter $\xi$
                   }%
\vskip15pt

\bigskip

\section*{ATTRACTION INFLUENCE ON THE SMALL OSCILLATION SPECTRUM}
In the majority of the papers devoted to
a non-ideal Bose-gas, as in~\cite{B}, it is supposed that
the inter-particle interaction is of repulsive type.
It is of interest to investigate the influence
of the inter-particle long-distance attraction on
the elementary excitation spectrum.

 Toward this end we calculate the spectrum for
 the interaction potential
\begin{equation}
\label{eq11}
U\left( {\mathbf{r}} \right) =
\left\{
\begin{array}{ll}
U_m,      & r \le a, \\
-C/r^6,   & r > a,
\end{array}
\right.
\end{equation}
\noindent where $C$ is positive.
This potential accounts for a Van der Waals
long-distance attraction of particles.
The dispersion law for potential~(\ref{eq11})
can be expressed as
\begin{equation}
\label{eq12} \omega ^{2} = \omega_{0} ^{2}
\left( f(z)-\alpha g(z) \right),
\end{equation}
\noindent with a dimensionless parameter $\alpha=C/U_{m}a^6$
and an auxiliary function
\begin{eqnarray}
\label{eq13}
g(z)&=&\frac{z}{4}
\left[ \left(1-\frac{z^2}{6}\right)\sin z
+ \frac{z}{3} \left(1-\frac{z^2}{2}\right) \cos z
\right. \\
\nonumber
&&\left.\quad\quad\quad
 + \frac{z^4}{6} \left(\frac{\pi}{2} - \mathrm{Si} (z)\right) \right],
\end{eqnarray}
\noindent ($Si\left( {z} \right)$ stands for integral sine)
describes the impact of the attractive part of the inter-particle
potential on the elementary excitation spectrum.
This function is shown in Fig.4.
Function $g(z)$ is positive for wave numbers $0 < z < 2.45$
and negative for $2.45 < z < 5.40$.
For $z > 5.40$ function $g\left( {z} \right)$
is positive and rises sharply.

\bigskip
\begin{center} \noindent
\epsfxsize=0.8\columnwidth\epsffile{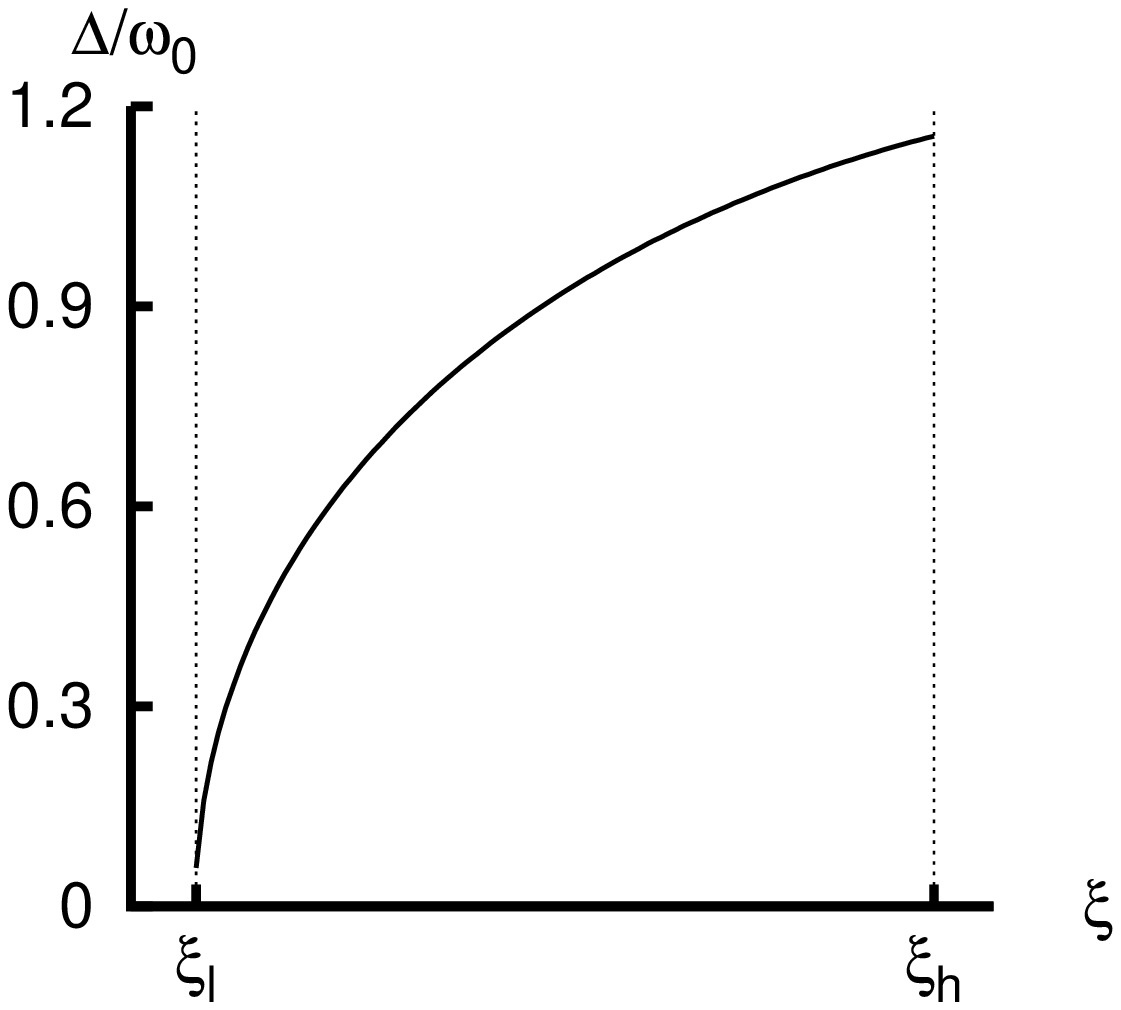}
\end{center}
\vskip-3mm\noindent{\footnotesize
Fig.3  Roton gap versus parameter $\xi$.
                   }%
\vskip15pt

One may notice that the interval $2.45 < z < 5.40$,
where function~(\ref{eq13}) is negative,
overlaps with the domain of the roton minimum existence.
It follows that the attraction leads to the growth of the excitation energy
in the domain of the roton minimum existence and smears this minimum.
In the range of small wave numbers,
and also for very large $z > 5.40$, attraction leads
to lowering the excitation energy.
Thus the long-distance atomic attraction results in
smoothing of the dispersion curve and in narrowing the range
of the existence of excitation spectrum, which has extrema
(rotons and maxons).
Attraction reduces the upper bound $\xi_{h}$
and increases the lower one $\xi_{l}$,
narrowing the range
of the existence of a Landau like excitation spectrum
$\delta \xi \equiv \xi_{h} - \xi_{l}$.
Dependence of $\delta \xi$ for a spectrum with roton minimum
on
attraction strength $\alpha = C \left/ U_{m}a^{6} \right.$,
is shown in Fig.5.

Notice that experimental spectrum of quasi-particle excitations
in the superfluid Helium at large momentum
 $(p > 3$\r{A}$^{-1})$ increases slowly such that
 the derivative of $\omega(p)$ is close to zero at the boundary of the
 spectrum~\cite{HG}.
 This behavior can be understood with help
 of a long-range attraction.
Recall that function $g(z)$ is positive and increasing for $z>5,4$.
Thereby, due to~(\ref{eq12}), it opposes the dispersion curve growing
at large momentum.

One should draw attention to smallness of the attraction
part of the potential for Helium --- it is the smallest among
the rare gases.
Namely, the depth of te potential well for Helium
 $\varepsilon = - 10.5$K, for Neon $\varepsilon = - 35.9$K,
 for Argon $\varepsilon = - 121$K,
 for Krypton $\varepsilon = - 173$K.
 It seems that due to this fact the maxon-roton
 part of the dispersion curve in liquid Helium
 is the most pronounced, not only in a superfluid phase,
 but in a normal phase as well.
 Notice a feature of the superfluid phase:
 the lifetime of the maxon-roton excitations are much larger
 than that in normal liquids.

\bigskip
\begin{center} \noindent
\epsfxsize=0.8\columnwidth\epsffile{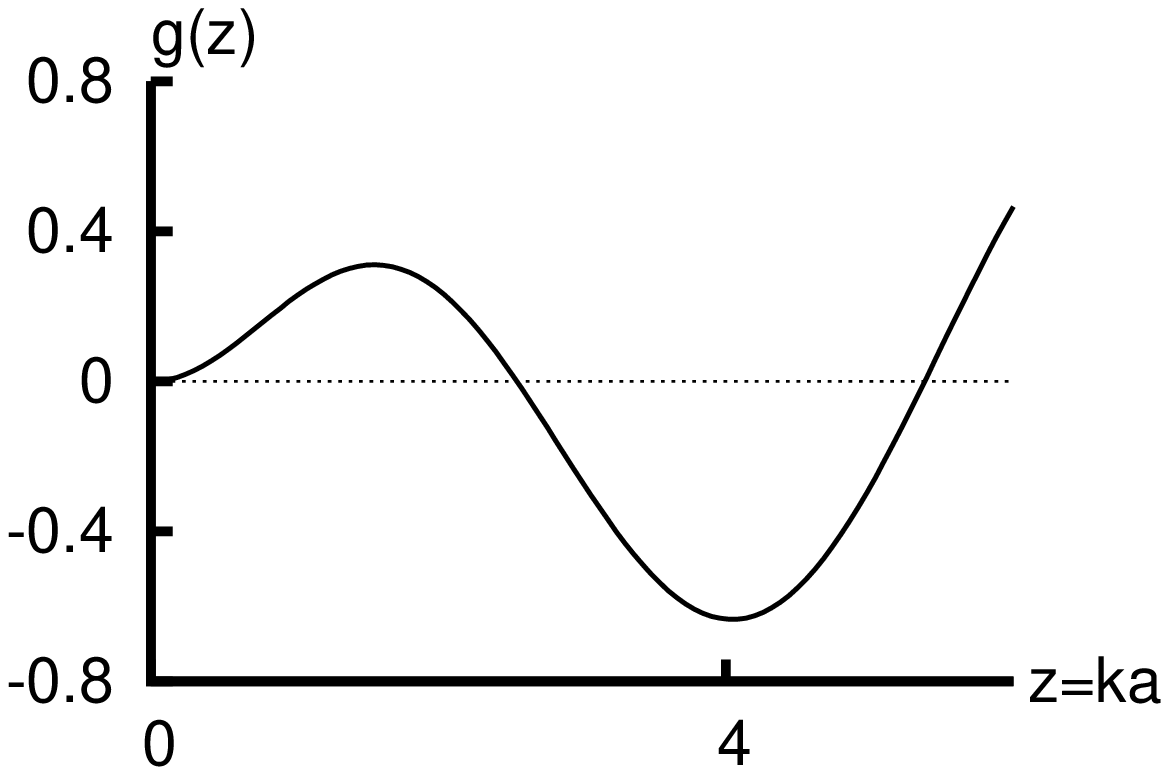}
\end{center}
\vskip-3mm\noindent{\footnotesize
Fig.4. Function $g\left( {z} \right)$ in the domain of $z$ of physical
interest.
                   }%
\vskip15pt

\bigskip
\begin{center} \noindent
\epsfxsize=0.8\columnwidth\epsffile{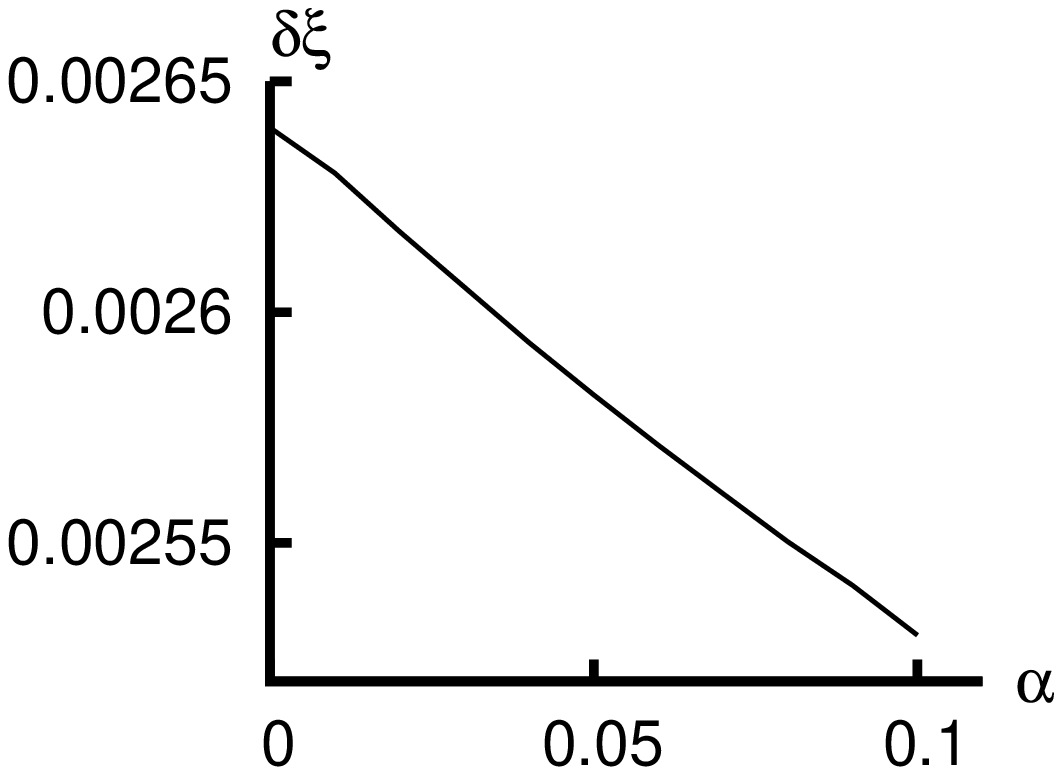}
\end{center}
\vskip-3mm\noindent{\footnotesize
Fig.5. Range $\delta \xi$ as a function of
the interaction strength $\alpha $.
                   }%
\vskip15pt


\section*{CONCLUSIONS}

It is shown that the Gross-Pitaevskii equation with
account for the non-local effects, caused by the finiteness of
the inter-particle interaction range, can describe
the short-wave excitation spectrum, for which the wave length
is of inter-particle interaction range order.

For a Bose-system the spectrum
for arbitrary wave number is obtained;
it has a shape of the Landau spectrum of a superfluid Helium.
We estimated the range of parameters where
such a spectrum exists.
Computation shows the roton minimum in the dispersion
curve is caused by the finiteness of the inter-particle interaction range;
the wave length of this minimum is close to the repulsion range of
the interaction.
The influence of the long-distance inter-particle attraction
on the dispersion curve is studied.
It is shown that it results in
smoothing of the excitation spectrum and in narrowing the range
of the existence of a maxon-roton type curve.
It is noticed that the attraction part of the potential
for the Helium atoms is small in comparison
with that for other atoms; it seems that this fact
induces a pronounced maxon-roton part of the liquid Helium
dispersion curve.

\rezume{%
КОРОТКОХВИЛЬОВІ ЗБУДЖЕННЯ В
\\НЕЛОКАЛЬНІЙ МОДЕЛІ ГРОССА-ПІТАЄВСЬКОГО}
{%
А.П.Івашин, Ю.М.Полуектов} {Показано, що нелокальна форма рівняння
Гросса-Пітаєвського дозволяє описати не тільки довгохвильові, але
й короткохвильові збудження в системах з бозе-конденсатом. При
певних значеннях параметрів, спектр збуджень бозе-конденсату
нагадує спектр Ландау квазічастинкових збуджень у надплинному
гелії з ротонним мінімумом. Довжина хвилі збуджень, при якій існує
ротонний мінімум, близька до радіуса міжчастинкової взаємодії.
Показано, що урахування міжчастинкового притягання звужує область
існування спектра з ротонним мінімумом.}

\rezume{%
КОРОТКОВОЛНОВЫЕ ВОЗМУЩЕНИЯ В \\НЕЛОКАЛЬНОЙ МОДЕЛИ ГРОССА-
ПИТАЕВСКОГО}
{%
А.П.Ивашин, Ю.М.Полуэктов} {Показано, что нелокальная форма
уравнения Гросса-Питаевского позволяет описать не только
длинноволновые, но и коротковолновые возбуждения в системах с
бозе-конденсатом. При определенных значениях параметров, спектр
возбуждений бозе-конденсата напоминает спектр Ландау
квазичастичных возбуждений в сверхтекучем гелии с ротонным
минимумом. Длина волны возбуждений, при которой существует
ротонный минимум, близка к радиусу межчастичного взаимодействия.
Показано, что учет межчастичного притяжения сужает область
существования спектра с ротонным минимумом.}

\end{multicols}

\begin{thebibliography}{99}
\bibitem{G}
E.P. Gross,  Nuovo Cimento {\bf 20}, 454 (1961).

\bibitem{P}
L.P. Pitaevskii,  JETP {\bf40}, 646 (1961).

\bibitem{LP}
E.M. Lifshits, L.P Pitaevskii, Statistical
Physics. Part.2 (Nauka, Moscow, 1978).

\bibitem{B}
N.N. Bogolyubov,  J. Phys. {\bf9}, 23 (1947).

\bibitem{A}
I.N. Adamenko , K.E. Nemchenko , I.V. Tanatarov,  Phys. Rev. B
{\bf67}, 104513 (2003).

\bibitem{EP}
E.A. Pashitskii,  Fiz. Nizk. Temp.  {\bf25}, N2, 115 (1999).

\bibitem{H}
 I.M. Khalatnikov, The theory of superfluidity (Nauka, Moscow, 1971).

\bibitem{HG}
H.R. Glyde, Phys. Rev. B {\bf45}, 7321 (1992).


\begin{flushright}
{\footnotesize Received 18.09.09}
\end{flushright}
\end{thebibliography}
\end{document}